\documentstyle[11pt,Fan]{article}
\begin{document}
\lefthead{FAN ET AL.}
\righthead{SDSS QUASARS}

\submitted{The Astronomical Journal, in press (July 1999)}

\title{High-Redshift Quasars Found in Sloan Digital
Sky Survey Commissioning Data$^1$}

\author{Xiaohui Fan\altaffilmark{\ref{Princeton}},
Michael A. Strauss\altaffilmark{\ref{Princeton}},
Donald P. Schneider\altaffilmark{\ref{PennState}}, 
James E. Gunn\altaffilmark{\ref{Princeton}}, 
Robert H. Lupton\altaffilmark{\ref{Princeton}}, 
Brian Yanny\altaffilmark{\ref{Fermilab}}, 
Scott F. Anderson\altaffilmark{\ref{Washington}}, 
John E. Anderson, Jr.\altaffilmark{\ref{Fermilab}},
James Annis\altaffilmark{\ref{Fermilab}},
Neta A. Bahcall\altaffilmark{\ref{Princeton}},
J. A. Bakken\altaffilmark{\ref{Fermilab}},
Steven Bastian\altaffilmark{\ref{Fermilab}},
Eileen Berman\altaffilmark{\ref{Fermilab}},
William N. Boroski\altaffilmark{\ref{Fermilab}},
Charlie Briegel\altaffilmark{\ref{Fermilab}},
John W. Briggs\altaffilmark{\ref{Yerkes}},
J. Brinkmann\altaffilmark{\ref{APO}}, 
Michael A. Carr\altaffilmark{\ref{Princeton}}, 
Patrick L. Colestock\altaffilmark{\ref{Fermilab}},
A. J. Connolly\altaffilmark{\ref{Pittsburgh}},
J. H. Crocker\altaffilmark{\ref{JHU}},
Istv\'an Csabai\altaffilmark{\ref{JHU},\ref{Eotvos}},
Paul C. Czarapata\altaffilmark{\ref{Fermilab}},
John Eric Davis\altaffilmark{\ref{APO}},
Mamoru Doi\altaffilmark{\ref{UTokyo}},
Brian R. Elms\altaffilmark{\ref{Princeton},\ref{NAOJapan}}
Michael L. Evans\altaffilmark{\ref{Washington}},
Glenn R. Federwitz\altaffilmark{\ref{Fermilab}},
Joshua A. Frieman\altaffilmark{\ref{Fermilab},\ref{Chicago}},
Masataka Fukugita\altaffilmark{\ref{CosmicRay},\ref{IAS}},
Vijay K. Gurbani\altaffilmark{\ref{Fermilab},\ref{IIT}},
Frederick H. Harris\altaffilmark{\ref{Flagstaff}},
Timothy M. Heckman\altaffilmark{\ref{JHU}},
G. S. Hennessy\altaffilmark{\ref{USNO}},
Robert B. Hindsley\altaffilmark{\ref{USNO}},
Donald J. Holmgren\altaffilmark{\ref{Fermilab}},
Charles Hull\altaffilmark{\ref{OCIW}},
Shin-Ichi Ichikawa\altaffilmark{\ref{NAOJapan}},
Takashi Ichikawa\altaffilmark{\ref{Tohoku}},
\v{Z}eljko Ivezi\'{c}\altaffilmark{\ref{Princeton}},
Stephen Kent\altaffilmark{\ref{Fermilab}},
G. R. Knapp\altaffilmark{\ref{Princeton}},
Richard G. Kron\altaffilmark{\ref{Chicago},\ref{Fermilab}},
D.Q. Lamb\altaffilmark{\ref{Chicago}},
R. French Leger\altaffilmark{\ref{Washington}},
Siriluk Limmongkol\altaffilmark{\ref{Washington}},
Carl Lindenmeyer\altaffilmark{\ref{Fermilab}},
Daniel C. Long\altaffilmark{\ref{APO}},
Jon Loveday\altaffilmark{\ref{Chicago}},
Bryan MacKinnon\altaffilmark{\ref{Fermilab},\ref{MerrillLynch}},
Edward J. Mannery\altaffilmark{\ref{Washington}},
P. M. Mantsch\altaffilmark{\ref{Fermilab}},
Bruce Margon\altaffilmark{\ref{Washington}},
Timothy A. McKay\altaffilmark{\ref{Michigan}},
Jeffrey A. Munn\altaffilmark{\ref{Flagstaff}},
Thomas Nash\altaffilmark{\ref{Fermilab}},
Heidi Jo Newberg\altaffilmark{\ref{Fermilab}},
R. C. Nichol\altaffilmark{\ref{CMU}},
Tom Nicinski\altaffilmark{\ref{Fermilab},\ref{Lucent}},
Sadanori Okamura\altaffilmark{\ref{UTokyo}},
Jeremiah P. Ostriker\altaffilmark{\ref{Princeton}},
Russell Owen\altaffilmark{\ref{Washington}},
A. George Pauls\altaffilmark{\ref{Princeton}},
John Peoples\altaffilmark{\ref{Fermilab}},
Donald Petravick\altaffilmark{\ref{Fermilab}},
Jeffrey R. Pier\altaffilmark{\ref{Flagstaff}},
Ruth Pordes\altaffilmark{\ref{Fermilab}},
Angela Prosapio\altaffilmark{\ref{Fermilab}},
Ron Rechenmacher\altaffilmark{\ref{Fermilab}},
Gordon T. Richards\altaffilmark{\ref{Chicago}},
Michael W. Richmond\altaffilmark{\ref{Rochester}},
Claudio H. Rivetta\altaffilmark{\ref{Fermilab}},
Constance M. Rockosi\altaffilmark{\ref{Chicago}}, 
Dale Sandford\altaffilmark{\ref{Yerkes}},
Gary Sergey\altaffilmark{\ref{Fermilab}},
Maki Sekiguchi\altaffilmark{\ref{CosmicRay}},
Kazuhiro Shimasaku\altaffilmark{\ref{UTokyo}},
Walter A. Siegmund\altaffilmark{\ref{Washington}},
J. Allyn Smith\altaffilmark{\ref{Michigan}},
Chris Stoughton\altaffilmark{\ref{Fermilab}},
Alexander S. Szalay\altaffilmark{\ref{JHU}},
Gyula P. Szokoly\altaffilmark{\ref{JHU}},
Douglas L. Tucker\altaffilmark{\ref{Fermilab}},
Michael S. Vogeley\altaffilmark{\ref{Princeton}},
Patrick Waddell\altaffilmark{\ref{Washington}},
Shu-i Wang\altaffilmark{\ref{Chicago}},
David H. Weinberg\altaffilmark{\ref{Ohio}}, 
Naoki Yasuda\altaffilmark{\ref{NAOJapan}}, and
Donald G. York\altaffilmark{\ref{Chicago}} 
}

\altaffiltext{1}{Based on observations obtained with the
Sloan Digital Sky Survey, and with the Apache Point Observatory
3.5-meter telescope, which is owned and operated by the Astrophysical
Research Consortium}
\newcounter{address}
\setcounter{address}{2}
\altaffiltext{\theaddress}{Princeton University Observatory, Princeton, NJ 08544
\label{Princeton}}
\addtocounter{address}{1}
\altaffiltext{\theaddress}{Department of Astronomy and Astrophysics,
The Pennsylvania State University,
University Park, PA 16802
\label{PennState}}
\addtocounter{address}{1}
\altaffiltext{\theaddress}{Fermi National Accelerator Laboratory, P.O. Box 500,
Batavia, IL 60510
\label{Fermilab}}
\addtocounter{address}{1}
\altaffiltext{\theaddress}{University of Washington, Department of Astronomy,
Box 351580, Seattle, WA 98195
\label{Washington}}
\addtocounter{address}{1}
\altaffiltext{\theaddress}{Yerkes Observatory, University of Chicago, 
     373 W. Geneva St. Williams Bay, WI 53191
\label{Yerkes}}
\addtocounter{address}{1}
\altaffiltext{\theaddress}{Apache Point Observatory, P.O. Box 59,
Sunspot, NM 88349-0059
\label{APO}}
\addtocounter{address}{1}
\altaffiltext{\theaddress}{Department of Physics and Astronomy
          University of Pittsburgh
          Pittsburgh PA 15260
\label{Pittsburgh}}
\addtocounter{address}{1}
\altaffiltext{\theaddress}{
Department of Physics and Astronomy, The Johns Hopkins University,
   3701 San Martin Drive, Baltimore, MD 21218, USA
\label{JHU}
}
\addtocounter{address}{1}
\altaffiltext{\theaddress}{Department of Physics of Complex Systems,
E\"otv\"os University,
   P\'azm\'any P\'eter s\'et\'any 1/A, Budapest, H-1117, Hungary
\label{Eotvos}
}
\addtocounter{address}{1}
\altaffiltext{\theaddress}{Department of Astronomy and Research Center for the Early Universe,
	School of Science, University of Tokyo, Hongo, Bunkyo,
Tokyo, 113-0033 Japan
\label{UTokyo}}
\addtocounter{address}{1}
\altaffiltext{\theaddress}{National Astronomical Observatory, 2-21-1, Osawa, Mitaka,
Tokyo 181-8588, Japan
\label{NAOJapan}}
\addtocounter{address}{1}
\altaffiltext{\theaddress}{University of Chicago, Astronomy \& Astrophysics
Center, 5640 S. Ellis Ave., Chicago, IL 60637
\label{Chicago}}
\addtocounter{address}{1}
\altaffiltext{\theaddress}{Institute for Cosmic Ray Research, University of
Tokyo, Midori, Tanashi, Tokyo 188-8502, Japan
\label{CosmicRay}}
\addtocounter{address}{1}
\altaffiltext{\theaddress}{Institute for Advanced Study, Olden Lane,
Princeton, NJ 08540
\label{IAS}}
\addtocounter{address}{1}
\altaffiltext{\theaddress}{Illinois Institute of Technology, 3300 S. Federal Street
Chicago, Il 60616
\label{IIT}}
\addtocounter{address}{1}
\altaffiltext{\theaddress}{U.S. Naval Observatory, Flagstaff Station, 
P.O. Box 1149, 
Flagstaff, AZ  86002-1149
\label{Flagstaff}}
\addtocounter{address}{1}
\altaffiltext{\theaddress}{U.S. Naval Observatory, 
3450 Massachusetts Ave., NW, 
Washington, DC  20392-5420
\label{USNO}}
\addtocounter{address}{1}
\altaffiltext{\theaddress}{The Observatories of the Carnegie Institution of 
Washington, 813 Santa Barbara St, Pasadena, CA 91101
\label{OCIW}}
\addtocounter{address}{1}
\altaffiltext{\theaddress}{Astronomical Institute,
Tohoku University,
Aoba, Sendai 980-8578
Japan
\label{Tohoku}}
\addtocounter{address}{1}
\altaffiltext{\theaddress}{Merrill Lynch, 
1-1-3 Otemachi, Chiyoda-ku, Tokyo 100, Japan
\label{MerrillLynch}}
\addtocounter{address}{1}
\altaffiltext{\theaddress}{University of Michigan, Department of Physics,
	500 East University, Ann Arbor, MI 48109
\label{Michigan}}
\addtocounter{address}{1}
\altaffiltext{\theaddress}{Dept. of Physics, Carnegie Mellon University,
     5000 Forbes Ave., Pittsburgh, PA-15232
\label{CMU}}
\addtocounter{address}{1}
\altaffiltext{\theaddress}{Lucent Technologies, 2000 N Naperville Rd,
Naperville, IL 60566
\label{Lucent}} 
\addtocounter{address}{1}
\altaffiltext{\theaddress} {Physics Department,
            Rochester Institute of Technology,
            85 Lomb Memorial Drive,
            Rochester, NY 14623-5603
\label{Rochester}}
\addtocounter{address}{1}
\altaffiltext{\theaddress}{Ohio State University, Dept.~of Astronomy, 174
W. 18th Ave., Columbus, OH 43210
\label{Ohio}}
\begin{abstract}
We present photometric and spectroscopic
observations of 15 high-redshift quasars ($z > 3.6$)  
discovered from $\sim 140$ deg$^2$ of five-color ($u'g'r'i'z'$)
imaging data taken by the Sloan Digital Sky Survey (SDSS)
during its commissioning phase. 
The quasars are selected by their distinctive colors in SDSS
multicolor space.
Four of the quasars have redshifts higher than 4.6 ($z=4.63$, 4.75,
4.90 and 5.00, the latter being the highest redshift quasar yet
known).  In addition, two previously known $z > 4$ objects were
recovered 
from the data.  The quasars all have $i^* < 20$ and have luminosities
comparable to that of 3C$\,$273.  The spectra of the quasars have similar
features (strong, broad emission lines and substantial absorption
blueward of the Ly$\alpha$ emission line) seen in previously known
high-redshift quasars.  Although the photometric accuracy and image
quality fail to meet the final survey requirements, our success rate
for identifying high-redshift quasars (17 quasars from 27 candidates)
is much higher than that of previous multicolor surveys.  However, the
numbers of high-redshift quasars found is in close accord with the
number density inferred from previous surveys. 
\end{abstract}

\section{Introduction}

The space density of optically luminous quasars peaks at $z \sim 3$, and
declines towards higher redshift 
(\cite{Osmer82}, \cite{Warren94}, \cite{SSG95},
\cite{Kennefick95}, \cite{HawkinsVeron96}). 
The presence of quasars at high redshift provides useful 
constraints on cosmological models
(\cite{ER88}, \cite{Turner91}, \cite{EL95}).
The evolution of the quasar population at high redshift, and
its relation with the evolution of the star forming rate and 
the intergalactic ionizing background, are important input to our
understanding of the
formation and early evolution of galaxies
(e.g. \cite{BT98}, \cite{Madau98}).
Meanwhile, the absorption spectra of luminous quasars at high redshift
are extremely important in understanding the intergalactic medium
(e.g., \cite{Rauch98}). 

Since the discovery of the first 
quasar at $z>4$ (\cite{Warren87}), more
than 60 quasars at $z>4$ have been found 
(\cite{WH90}, \cite{Schneider99}, \cite{Kennefick99}).  However, only
12 quasars at  $z > 4.5$ have been identified, of which only 
one (PC1247+3406, $z=4.90$, \cite{SSG91b}) has a redshift larger
than~4.74. Indeed, \cite{Fan99} shows that current data are consistent
with only one $z>4.5$ 
QSO per 50 deg$^2$ for $i'<20$.

Most known high-redshift quasars have been selected as outliers in
color space from the stellar locus using multicolor broadband
photometry (e.g., \cite{Warren87}, \cite{Kennefick95}, \cite{SL96}). 
At $z>4.5$, the major emission lines and absorption features of
quasars are redshifted into the red part of the optical band.
Searching for quasars at these redshifts poses three major
technical requirements:  a red-sensitive detector and filter system,
large sky coverage (because of their rarity), and fast, automated software to do accurate
photometry on the resulting large data set. 


The Sloan Digital Sky Survey (SDSS; 
\cite{GW95}, \cite{SDSS96}, \cite{York99}) will use a dedicated  2.5m telescope
to obtain CCD images to $\sim 23^m$ in five broad optical bands
($u'$, $g'$, $r'$, $i'$, $z'$, 
centered at 3540\AA, 4770\AA, 6230\AA, 7630\AA\ and 9130\AA,
\cite{F96}) over 10,000 deg$^2$ 
of the high Galactic latitude sky
centered approximately on the North Galactic Pole. 
The survey data processing software provides astrometric and photometric
calibrations, and finds and measures properties of all detected objects
in the data. Spectra of the brightest $10^6$ galaxies and $1.5 \times
10^5$ quasar candidates will be obtained
on the same telescope. 
The SDSS will select quasar candidates from the much more numerous
stars by their distinctive colors, 
and will generate a large and uniformly-selected sample covering
 redshifts ranging from $z \sim 0$ to $z>5$.
Because of the large sky coverage and the inclusion of the
near-infrared $z'$ band, the SDSS is capable of identifying large
numbers of quasars at $z>4.5$ (Fan 1999).

The SDSS achieved first light in imaging mode in May 1998.
By the end of 1998, it had acquired several hundred deg$^2$ of imaging data
in its commissioning phase.
Using the ARC 3.5m telescope, we have obtained spectra of a number of
high-redshift quasar candidates produced by these data. 
In this paper, we report the discovery of 15 high-redshift quasars at
$z>3.6$, including four quasars at $z> 4.6$, 
identified from $\sim 140$ deg$^2$ of SDSS imaging data, 
1.4\% of the total survey area of the SDSS.
In \S 2, we discuss the photometric observations and data
reduction. In \S 3, we discuss the color selection of high-redshift
quasar candidates. The
spectroscopic observations of these objects are presented in \S 4. 
In \S 5, we discuss the spectral
properties of the quasars.
We briefly discuss the success rate and the quasar number counts
we find from these data, and compare the number counts with predictions
based on previous results on quasar evolution in \S 6. 

\section{Photometric Observations}

The SDSS telescope is a dedicated 2.5m altitude-azimuth instrument at
Apache Point, New Mexico, 
with a wide, essentially distortion-free field.
It has a modified $f/5.0$ Ritchey-Chretien optical design with a large
secondary and two corrector lenses below the primary mirror
(\cite{Siegmund99}, \cite{SDSS96}). 
Spectroscopy and imaging require separate second correctors.
The imaging corrector forms the base structural element for 
the imaging camera. 

The imaging camera (\cite{Gunnetal}) consists of a mosaic
of 30 imaging $2048 \times 2048$ SITe CCDs with 24$\mu$m pixels
subtending $0.4''$ on the sky.  
The CCDs are arranged in six dewars (six columns) containing 5 CCDs each,
mounted to the corrector and observing the sky through five broad-band
filters.  The five filters ($u'$, $g'$, $r'$, $i'$, and $z'$) cover
the entire optical band from the atmospheric cutoff in the blue to
the silicon sensitivity cutoff in the red. 
The $u'$ band, with the bulk of its response shortward of the
Balmer discontinuity,  
is especially important for the selection of low-redshift
quasars, while the reddest band, $z'$, is critical for quasars at 
$z>4.5$, where the Ly$\alpha$ emission line enters the $i'$ band.
The photometric data are taken in time-delay and integrate (TDI, or
``drift-scan'') mode at sidereal rate; a given point on the sky passes
through each of the five filters in succession.  
The total integration time per filter is 54.1 seconds, and
the expected survey depth (5 $\sigma$ detection for point
sources at $1''$ seeing) is
22.3, 23.3, 23.1, 22.5 and 20.8 in $u'$, $g'$, $r'$, $i'$, and $z'$,
respectively (but see below). 
The imaging CCDs saturate at about $\rm 14^m$, so to calibrate these
data with existing astrometric catalogs, the camera contains an additional 22
CCDs, each $2048 \times 400$ pixels, with neutral density filters that
saturate only at $\rm 8^m$ (\cite{Gunnetal}, \cite{ASTROM}).  The data
are read from the CCDs, and onto disk and tape using a data
acquisition system described in \cite{Petravick94} and
\cite{Petravick99}. 
%

The SDSS photometric system 
is based on the AB$_{\nu}$ system:

\begin{equation}
m = -2.5 \log \frac{\int d(\log \nu) f_{\nu} S_{\nu}}
                {\int d(\log \nu)  S_{\nu}} - 48.60, 
\end{equation}

\noindent
with the zeropoints defined by the spectrophotometric observations
of the \cite{OG83} standard stars (\cite{F96}).
The photometric calibration will be carried out with  a separate $20''$ 
photometric telescope equipped with a single-CCD camera and the SDSS
filters (\cite{Uomoto99}).
The photometric telescope will continually observe photometric standard stars to 
measure the photometric zeropoint and atmospheric extinction.
However, the data used in this paper were taken before the photometric
telescope was
commissioned; we thus calibrated the data by observing secondary  
standards in the survey area (cf.\ \cite{Smith98}, \cite{Tucker98})
using a (now decommissioned) $24''$ telescope at Apache Point and the US Naval
Observatory's $40''$ telescope. 
Moreover, the SDSS primary standard star network had not
been completely established when these data were taken. 
Therefore, the absolute photometric calibration is only accurate to
5--10\% (although the relative photometry is considerably better).
Thus in this paper, we will denote the preliminary SDSS magnitudes we
have measured as $u^*, g^*, r^*, i^*$ and $z^*$,
rather than the notation $u'$, $g'$, $r'$, $i'$, and $z'$ that will be
used for the final SDSS photometric system (and is used in this paper
to refer to the SDSS filters themselves). 

The SDSS data used in this paper were acquired with the telescope
parked on the meridian, at the Celestial Equator.  Most of the quasar candidates are
from data taken in 19 Sept 1998 (SDSS run 94).  This observation covers the
Northern strip of the equatorial scan  
(two interleaved scan {\em strips} are needed to fill in the
gaps between different columns of CCDs and form  a filled
{\em stripe} 2.5 degrees wide, see \cite{Gunnetal}).
The scan is 6 hours long along the Celestial Equator.  The raw data total roughly 120 Gbyte.
The scan covers the RA range of $\sim 22^h - 4^h$, with a total area
of $\sim 110$ deg$^2$ at high Galactic latitude ($|b| > 40^\circ$).
The night was photometric, with seeing varying from $1.2''$
to $2.0''$.
Because of low-level astigmatism in the 2.5m telescope optics during the 
early commissioning period, the width of
the Point Spread Function (PSF) is also a function of the position
of the star on the focal plane.
This is an important term in the error budget of the PSF photometry.
In addition, one quasar candidate was taken from data obtained on 21
Sept 1998 (SDSS run 109), which covered the Southern strip of the equatorial scan
from RA = 2$^h$ to 4$^h$, with a total area of $\sim 30$ deg$^2$.  The
seeing in this scan varied from $1.1''$ 
to $1.4''$. 

These data  are
processed by a series of automated pipelines to carry out
astrometric and photometric measurements.
The astrometric pipeline (\cite{ASTROM}) calibrates the positions of stars
with reference to the astrometric standards, and establishes an astrometric
solution for each of the imaging CCDs, giving accurate relative
astrometry between colors.  These data are analyzed with a pipeline
(\cite{Tucker98}) that measures the magnitudes of secondary standard 
stars, and sets the photometric zeropoint of the imaging camera, while
producing a measure of extinction on each night. 
The photometric pipeline (PHOTO, \cite{Lupton99b}) reduces the data from
the imaging camera and produces corrected images and object catalogs.
PHOTO carries out the following tasks:
\begin{enumerate} 
\item determines smoothly varying flat fields, bias vectors, sky
values, and PSFs along each scan line;
\item flat fields the data, interpolates over bad columns, and
removes cosmic rays;
\item finds objects in each color as statistically significant
peaks after smoothing with the PSF;
\item combines the data from five bands for each object;
\item deblends overlapped objects as described below;
\item measures the position, counts, size and shape of each detected object,
and 
\item classifies objects as being extended or consistent with point
sources, based on fitting simple models. 
\end{enumerate}
The final step is to write out calibrated object catalogs using
information passed from astrometric and photometric calibrations.
Flags indicating image processing problems (such as the presence of
saturated pixels, problems with the deblender, and the fact that a cosmic
ray has been removed), are also recorded in the output catalogs.	

PHOTO measures the PSF magnitude of each source by fitting
a PSF model of a 
double Gaussian. 
We will use this PSF magnitude throughout the paper.  
For objects that are undetected in a
given band, the PSF fit is constrained to the canonical center of the
object in that color, after offsetting according to the astrometric
solution.
This allows an unbiased measure of the flux of a source even in the
case when it is too faint to center up on it. 
Even though such a measurement is quite noisy, it is very important in
distinguishing between red stars and high-redshift quasars (\S 4).

Overlapping objects are deblended into their constituent parts
(``children") consistently in all five bands.  Peaks are found in the
image of a given object in each of the five bands, and the union of all the
peaks is used to define the children into which the object will be
deblended.  For each pixel, the deblender assigns weights to each
child based on the ansatz that astronomical objects possess a 
center of symmetry.  However, the final deblended children have no
such requirement on their symmetry.  The output of the deblender
is an image of each child in each filter, which are then passed to the
object measuring code. Details of the algorithm are given by
\cite{deblend}. 

The final object catalog of Run 94 includes a total of about 2 million
objects.  The limiting magnitudes are roughly 22.3, 22.6, 22.7, 22.4
and 20.5 in $u^*$, $g^*$, $r^*$, $i^*$, and $z^*$, respectively, as
determined from the cutoff in the observed number-magnitude relation.
This is somewhat brighter than the survey design due to the poor image
quality of this run, and the poor reflectivity of the primary mirror
(which has since been realuminized).  As mentioned above, the absolute
photometric calibration is accurate to $\sim 0.1$ mag, while the
relative photometric errors are better than 0.05 mag at $u^*< 20.7$,
$g^*<21.3$, $r^* < 20.8$, $i^*<20.1$, and $z^*<18.5$, respectively.
The error in the absolute photometric calibration has little effect on
the separation of quasars and stars in color space, as it shifts the
colors and magnitudes of all objects uniformly.

\section{Quasar Selection}

Figure 1 presents the color-color diagrams of one column (one-sixth of the
total) of the run 94 data for stellar sources at $i^* < 20$. 
A source is plotted in each panel only if it is detected in all three
of the relevant bands. 
Objects flagged as saturated in any band, lying on the bleed trail of
a saturated star, or overlapping the edge of the image boundary have
been rejected.  However, deblended objects are included in this
diagram; their distribution is 
no broader than that of isolated objects.  
The group of sources at $g^*-r^* \sim 1.5$ and $r^*-i^* \sim 0.7$ are
mostly compact galaxies at $i^* \sim 20$ (i.e., they appear not to be
point-like on the images); the star-galaxy classifier
was not fine-tuned at faint magnitudes when these data were processed.
Note the narrowness of the distributions of all the diagrams. 
We also plot the median tracks of quasar colors as a function of
redshift from the simulation of Fan (1999), and indicate the approximate
locations of low-redshift ($z< 2.5$) quasars, hot white dwarfs (WD) and
A stars from Fan (1999).

\begin{figure*}[t]
\plotfiddle{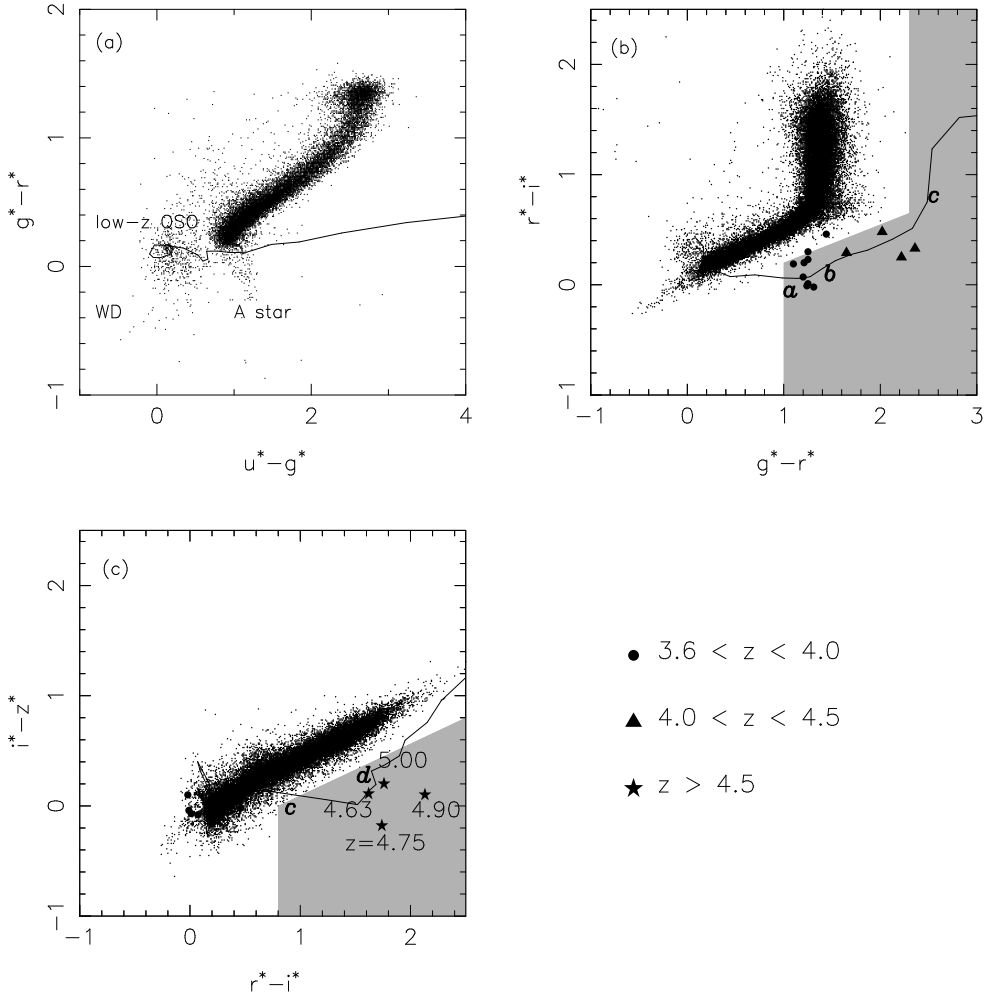}{5.20in}{0.}{70.}{70.}{-247}{-20}
\figurenum{1}

\protect\caption
{\footnotesize
 Color-color diagrams of stellar objects in 20 deg$^2$ of SDSS imaging data
in Run 94 at $i^* < 20$.
The shaded areas on the $g^*- r^*$ vs. $r^* - i^*$ and
the $r^* - i^*$ vs. $i^* - z^*$ diagrams represent the
selection criteria used to select $gri$ and $riz$ quasar
candidates.
The solid line is the median track of simulated quasar colors
as a function of redshift (adapted from Fan 1999).  The letters a, b,
c, and d indicate the positions on the locus of median color quasars
at $z = 3.6, 4.0, 4.6$, and 5.0, respectively.
Colors of the 17 confirmed SDSS quasars at $z>3.65$ are also plotted
on the diagrams.}
\label{fig:fig1}
\end{figure*}

Quasars are well-separated from the stellar locus at most redshifts in
the SDSS photometric system, and can be selected by their distinct
colors.  At $z<2.0$, quasar optical colors are dominated by the
power-law continuum, and the $u^* - g^*$ color of quasars remains
smaller than 0.5, considerably bluer than normal stars seen at high
latitudes.  Ly $\alpha$ emission enters the $u'$ band at $z\sim 2.0$,
making the $u^* - g^*$ color even bluer.  When Ly $\alpha$ emission
moves from $u'$ to $g'$ at $z \sim 2.5$, the $u^* - g^*$ color gets
redder.  For $z>2.5$, the absorption systems, first the Ly $\alpha$
forest, then the Lyman Limit Systems (LLSs), enter the $u'$ band.
They absorb a substantial fraction of the continuum radiation in $u'$
band, and the $u^* - g^*$ color reddens quickly with redshift.  At $z
\sim 2.8$, quasars have very similar colors to A stars in the SDSS system.

For $z > 3.6$, most quasars are undetectable in $u'$ due to the presence of
LLSs, which are optically thick to the continuum radiation from the quasar. 
Meanwhile, the absorption systems begin to dominate in the $g'$ band,
and Ly$\alpha$ emission moves to the $r'$ band. 
The $g^* - r^*$ color becomes increasingly red towards higher redshift,
while the $r^* - i^*$ color remains relatively small.
Quasars at $z>3.6$ are well-separated from the stellar locus in
the $g^* - r^*$ vs. $r^* - i^*$ diagram.
At $z>4.6$, quasars are well-separated  from the stellar locus in
the $r^* - i^*$ vs. $i^* - z^*$ diagram as the absorption features
move into the $r'$ band.
The first quasar at $z>4$ (\cite{Warren87}), 
as well as the highest redshift quasar previously
known (\cite{SSG91b}), were discovered using similar methods.
Recently, a similar technique has been  used to search for 
high-redshift galaxies, resulting in the
discovery of large scale clustering of Lyman break galaxies at
$z>3$ (\cite{Steidel98}) 
and the first discoveries  of galaxies  at $z>5$
(\cite{Spinrad98}, \cite{Weymann98}).

In the SDSS spectroscopic quasar survey, quasar candidates will be
selected by the target selection pipeline (cf. Newberg \& Yanny 1997).
At the time of writing, this pipeline has not been fine-tuned. 
Furthermore, the SDSS multi-fiber spectrograph has not yet been commissioned.
We therefore have carried out spectroscopic observations of only a small
number of objects.  We have applied simple color cuts
to select only quasars with colors consistent with having $z > 3.6$.



We carry out two separate cuts: 

1. $gri$ candidates: Quasar candidates with colors consistent with $z
> 3.6$, selected principally from their position  on the $g^* - r^*$
vs. $r^* - i^*$ diagram.  The full selection criteria are:

\begin{equation}
 \begin{array}{l}
	(a)\ i^* < 20 \\
	(b)\ u^* - g^* > 2.0 \mbox{ or } u^* > 22.3 \\
	(c)\ g^* - r^* > 1.0 \\
	(d)\ r^* - i^* < 0.2 + 0.42 (g^* - r^* -1.0) \mbox{ or } g^* - r^* >2.3 \\
	(e)\ i^* - z^* < 0.25
	\end{array}
\end{equation}

The region satisfied by criteria ($c$) and ($d$) is plotted as the
shaded area in Figure 1(b). 
Criterion ($b$) selects objects for which 
most of the flux in $u'$ is absorbed at $z>3.6$.
Criterion ($e$) provides the additional  constraint that
the continuum be relatively flat in the red part of the spectrum: this is
typical for quasars with power law continua; red stars and compact
red galaxies tend to have a much redder $i^* - z^*$. 

2. $riz$ candidates:  Quasar candidates with colors consistent with $z
> 4.6$, selected principally from their position on the  $r^* - i^*$ vs. $i^* - z^*$
diagram. The full selection criteria are:

\begin{equation}
\begin{array}{l}
	(a)\ i^* < 20.2  \\
	(b)\ u^* > 22.3 \\
	(c)\ g^* > 22.6 \\
	(d)\ r^* - i^* > 0.8 \\
	(e)\ i^* - z^* < 0.47 (r^* - i^* - 0.8) 
	\end{array}
\end{equation}

Quasars at $z>4.6$ have most of the flux absorbed in the $u'$ and
$g'$ bands; such objects are selected with criteria ($b$) and ($c$). 
  Criteria ($c$) and ($d$) are illustrated in Figure 1(c) as the
shaded region.  Needless to say, most of the $riz$ candidates are
undetected in $g$, and therefore do not show up in Figure 1(b). 

 We visually inspected the image data in all five filters for
all the candidates, paying special attention to those that were  
deblended from overlapping objects. We deleted from the candidate list
obvious  
problematic cases,  such as objects that are close to a saturated star
or big galaxy. 
The final candidate list includes 35 candidates;
25 are $gri$ candidates
at $i^* <20$, and $10$ are $riz$ candidates at $i^* < 20.2$. 
Eight candidates are deblended from nearby neighbors.
Two of the candidates, SDSSp J015339.61$-$001104.9 and SDSSp J010619.25+004823.4
(see Table 1; for naming convention, see \S 4), 
are the previously known quasars BRI0151--0025 ($z=4.2$) 
and BRI0103+0032 ($z=4.43$), found in a multicolor survey
(\cite{SL96}).  These two objects are the only quasars with 
$z>3.6$ in the area covered by Run 94 found in the NED
database\footnote{The NASA/IPAC Extragalactic Database (NED) is
operated by the Jet Propulsion Laboratory, 
California Institute of Technology, 
under contract with the National Aeronautics and Space Administration. }.
We matched the candidate
list against the FIRST (\cite{Becker95}) radio continuum catalog
(which covers this area of sky); only one of them has a radio
counterpart at 20 cm at the 1mJy level, namely SDSSp
J015339.61$-$001104.9, which is an unresolved source of 4.75 mJy.  We
have similarly cross-correlated the list against the ROSAT full-sky
pixel images; none of these objects were detected, implying a typical upper
limit of $2 \times 10^{-13} \rm\, erg\,cm^{-2}\,s^{-1}$ in the 0.1-2.4
keV band (W. Voges, private communication). 

\section{Spectroscopic Observations}

\begin{figure*}[t]
\plotfiddle{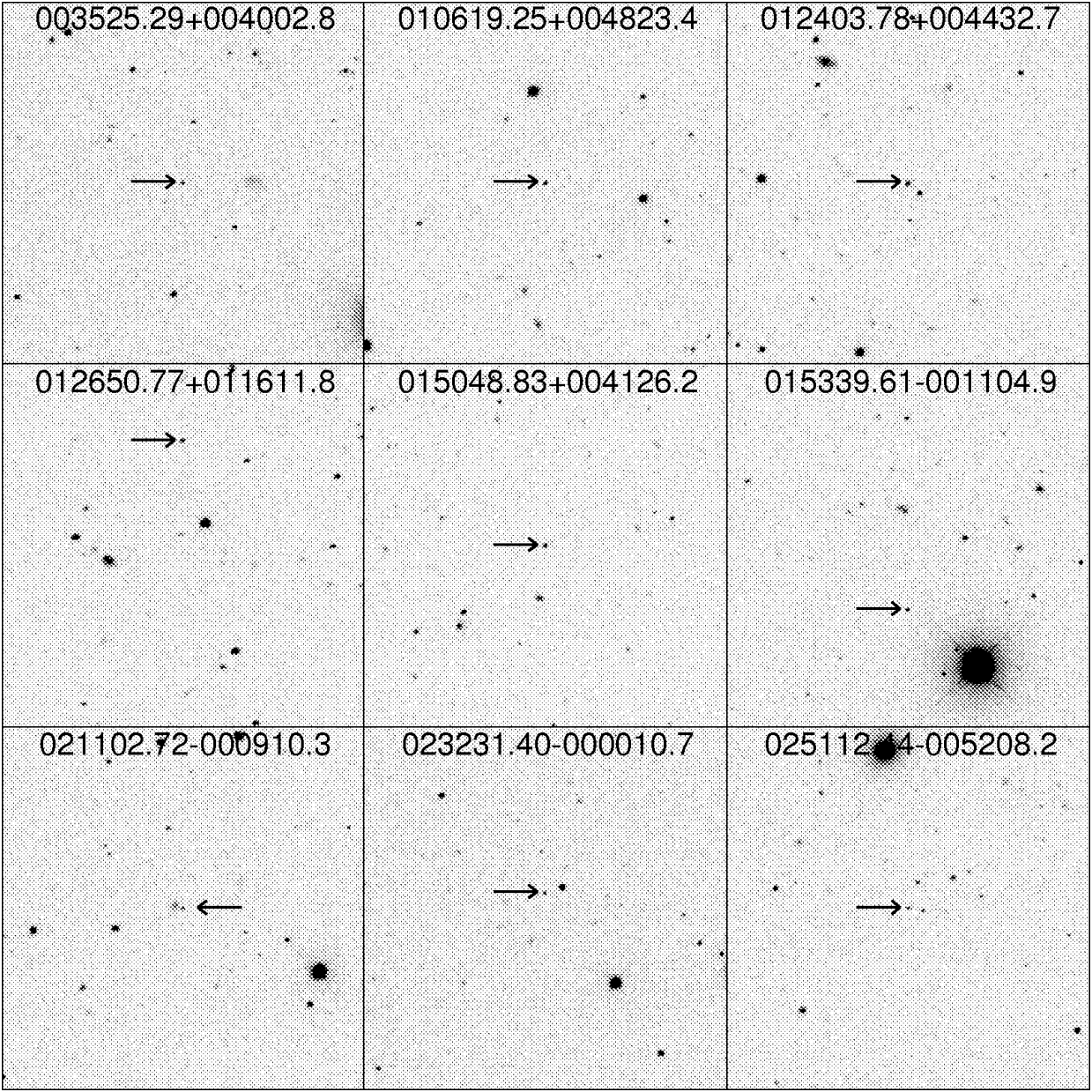}{7.25in}{0.}{80.}{80.}{-257}{-20}
\figurenum{2}
\protect\caption
{\footnotesize
Finding charts for the 17 SDSS quasars.
The data are $200'' \times 200''$ SDSS images in the $i'$ band (55 sec
exposure time).  North is up; East is to the left.}
\label{fig:fig2}
\end{figure*}

\begin{figure*}[t]
\plotfiddle{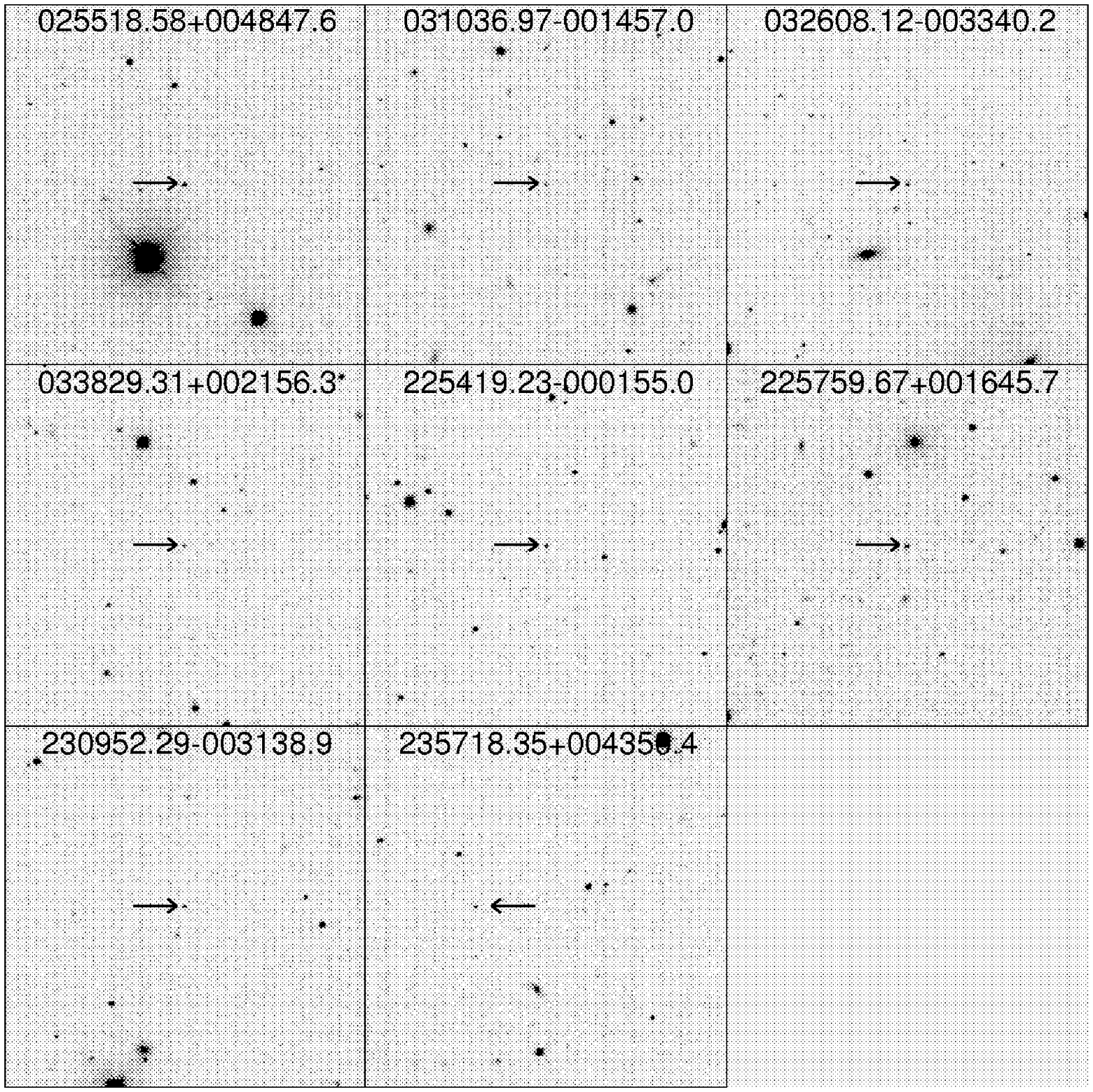}{7.25in}{0.}{80.}{80.}{-257}{-20}
\figurenum{2}
\protect\caption
{\footnotesize
Continued}
\label{fig:fig2}
\end{figure*}

Spectra of 24 high-redshift quasar candidates from Run 94, and one
from Run 109, were obtained with the ARC 3.5m telescope in
the Apache Point Observatory, using the Double Imaging Spectrograph (DIS)
during a number of nights in November and December of
1998.  One of the 24 objects observed (SDSSp J025518.58+004847.6)
does not satisfy the color selection criteria (eq.~2) with improved
photometric calibration, and although we do present its spectrum below, we
do not include it in the statistical analysis in \S 6. 

The DIS is a double spectrograph with a transition wavelength of 5350 \AA\
between the blue and red side.  The observations were taken with the
low resolution gratings, with a dispersion of 6.2 \AA\ pixel$^{-1}$ in
the blue side and 7.1 \AA\ pixel$^{-1}$ in the red side, and a
resolution of roughly 2 pixels.  The central wavelengths are 4400 \AA\
and 7700 \AA\ for the blue and red sides, respectively.  The final
spectrum covers the wavelength range of 4000 \AA\ to 10,000 \AA.  The
wavelength scale is calibrated with a cubic polynomial fit to lines
from an Ar-He-Ne lamp; the typical rms error of the fits is smaller
than 0.5 \AA.  Observations of the F subdwarf standards BD+17$^\circ$
4708 and HD 19445 (Oke \& Gunn 1983) provided flux calibration and
allowed removal of the atmospheric absorption bands.  However, most
of the candidates were not observed under photometric conditions, nor
were they observed with the slit oriented perpendicular to the horizon.
Exposure times ranged from 1200 seconds for an 18 mag candidate to 7200
seconds for the faintest candidates ($i^* \sim 20$).

The data were reduced using both standard software from the IRAF package
(by XF), and the procedures described in Schneider, Schmidt, \& Gunn
(1991a) (by DPS).  
We place the data on an absolute flux scale (to compensate for
the losses due to non-photometric conditions) by forcing
the synthetic $i^*$ magnitudes from the spectra to be the same
as the photometric measurements.
The two independent reductions produce consistent results.
Fifteen of the 25 candidates display the characteristics of previously
known high-redshift quasars: broad, strong emission lines, and
absorption due to the Ly$\alpha$ forest and Lyman Limit systems.  The
redshifts range from 3.66 to 5.00;  four of the $riz$ candidates are
quasars at redshifts 5.00, 4.90, 4.75 and 4.63. 

\begin{footnotesize}
\begin{deluxetable}{lcccccc}
\tablenum{1}
\tablecolumns{7}
\tablewidth{0pc}
\tablecaption{Positions and Photometry of SDSS High-redshift Quasars}
\tablehead
{
SDSS name & redshift & $u^*$ & $g^*$ & $r^*$ & $i^*$ & $z^*$
}
\startdata
SDSSp J003525.29+004002.8 & 4.75
& 24.12 $\pm$ 0.45   & 23.63 $\pm$   0.34 & 21.16 $\pm$  0.06 & 19.42 $\pm$  0.02 & 19.60 $\pm$  0.09 \\
SDSSp J010619.25+004823.4$^a$ & 4.43
& 23.86 $\pm$ 0.43   & 21.11 $\pm$  0.04 & 19.09 $\pm$  0.01 & 18.62 $\pm$  0.01 & 18.37 $\pm$  0.03 \\
SDSSp J012403.78+004432.7 & 3.81
& 22.73 $\pm$  0.20 & 19.18 $\pm$  0.01 & 17.87 $\pm$  0.01 & 17.89 $\pm$
 0.01 & 17.79 $\pm$  0.02 \\
SDSSp J012650.77+011611.8 & 3.66
& 22.76 $\pm$  0.28 & 20.44 $\pm$  0.02 & 19.20 $\pm$  0.01 & 19.22 $\pm$
 0.02 & 19.25 $\pm$  0.08 \\
SDSSp J015048.83+004126.2 & 3.67
& 23.27 $\pm$ 0.29  & 19.49 $\pm$  0.01 & 18.39 $\pm$  0.01 & 18.20 $\pm$
 0.01 & 18.14 $\pm$  0.02 \\
SDSSp J015339.61--001104.9$^b$ & 4.20
& 24.01 $\pm$ 0.43  & 21.14 $\pm$  0.04 & 18.92 $\pm$  0.01 & 18.67 $\pm$
 0.01 & 18.61 $\pm$  0.04 \\
SDSSp J021102.72--000910.3 & 4.90
& 24.04 $\pm$ 0.43  & 24.80 $\pm$  0.54 & 22.04 $\pm$  0.11 & 19.93 $\pm$  0.04 & 19.81 $\pm$  0.13 \\
SDSSp J023231.40--000010.7 & 3.81
& 24.87 $\pm$ 0.36  & 21.03 $\pm$  0.03 & 19.82 $\pm$  0.02 & 19.62 $\pm$
 0.03 & 19.78 $\pm$  0.12 \\
SDSSp J025112.44--005208.2 & 3.78
& 24.81 $\pm$ 0.45  & 20.67 $\pm$  0.02 & 19.42 $\pm$  0.01 & 19.41 $\pm$  0.02 & 19.48 $\pm$  0.10 \\
SDSSp J025518.58+004847.6$^c$ & 3.97
& 23.92 $\pm$ 0.44  & 20.53 $\pm$  0.03 & 19.09 $\pm$  0.01 & 18.63 $\pm$  0.01 & 18.65 $\pm$  0.04 \\
SDSSp J031036.97--001457.0 & 4.63
& 24.16 $\pm$ 0.34  & 23.41 $\pm$  0.20 & 21.51 $\pm$  0.06 & 19.89 $\pm$  0.02 & 19.77 $\pm$ 0.07 \\
SDSSp J032608.12--003340.2 & 4.16
& 23.25 $\pm$ 0.33  & 21.14 $\pm$  0.04 & 19.50 $\pm$  0.02 & 19.20 $\pm$  0.02 & 19.07 $\pm$  0.07 \\
SDSSp J033829.31+002156.3 & 5.00
& 24.27 $\pm$ 0.44  & 25.03 $\pm$  0.51 & 21.70 $\pm$  0.09 & 19.96 $\pm$  0.03 & 19.74 $\pm$  0.12 \\
SDSSp J225419.23--000155.0 & 3.68
& 23.31 $\pm$ 0.38  & 20.80 $\pm$  0.04 & 19.55 $\pm$  0.02 & 19.33 $\pm$  0.02 &
 19.27 $\pm$  0.10 \\
SDSSp J225759.67+001645.7 & 3.75
& 22.91 $\pm$ 0.31  & 20.11 $\pm$  0.02 & 18.92 $\pm$  0.01 & 18.84 $\pm$  0.02 &
 18.92 $\pm$  0.07 \\
SDSSp J230952.29--003138.9 & 3.95
& 24.06 $\pm$ 0.54  & 20.95 $\pm$  0.04 & 19.70 $\pm$  0.02 & 19.41 $\pm$  0.03 &
 19.42 $\pm$  0.10 \\
SDSSp J235718.35+004350.4 & 4.34
& 24.11 $\pm$ 0.53 & 22.43 $\pm$  0.14 & 20.08 $\pm$  0.03 & 19.75 $\pm$  0.04 &
 19.58 $\pm$  0.09 
\enddata
\tablenotetext{}{Asinh magnitudes (Lupton, Gunn \& Szalay 1999; see the Appendix) are
quoted; errors are statistical only.}

\tablenotetext{a}{This is the previously known quasar BRI0103+0032 (Storrie-Lombardi {\em et al.} 1996).}

\tablenotetext{b}{This is the previously known quasar BRI0151--0025 (Storrie-Lombardi {\em et al.} 1996).}

\tablenotetext{c}{This object is not part of the uniformly selected sample; it falls
just outside the color criteria listed in the text.}
\end{deluxetable}
\end{footnotesize}

Table 1 gives the position, SDSS photometry and redshift of each confirmed
SDSS quasar.
We also include in Table 1 the SDSS photometry of the two previously known
$z>4$ quasars in the survey region.
The naming convention for the SDSS sources is
SDSSp J$\,$HHMMSS.SS$\pm$DDMMSS.S, where ``p'' stands for the preliminary
SDSS astrometry (as the astrometric pipeline is still under
final commissioning), and the positions are expressed in J2000.0
coordinates. The preliminary SDSS astrometry is accurate to 
better than $0.2''$ rms in each coordinate.
The photometry is actually expressed in asinh magnitudes
(\cite{Luptitude}), which are identical to ordinary logarithmic
magnitudes in the high signal-to-noise ratio regime, but have much
better error properties at low signal-to-noise ratio, and indeed
allow one to express negative fluxes without simply quoting upper
limits.  They are described in detail in the Appendix. The errors
given are statistical in nature, and do not 
include systematic errors due to uncertainties in the photometric
zeropoint (of order 0.1 mag systematic in each band) and unmodeled
variations in the PSF (of order 0.05 mag in each band). 

The $z=4.90$ quasar (SDSSp J021102.72-000910.3)
is deblended from a blue $i^*=20$ galaxy $5''$ away.
This demonstrates that the photometric pipeline is able to deblend close
pairs and measure their properties accurately, even for a red point
source blended with a blue extended source. 

Finding charts of all objects in Table 1 are given in Figure 2.  The
images are $200'' \times 200''$ SDSS images in the $i'$ band. 

\begin{figure*}[t]
\plotfiddle{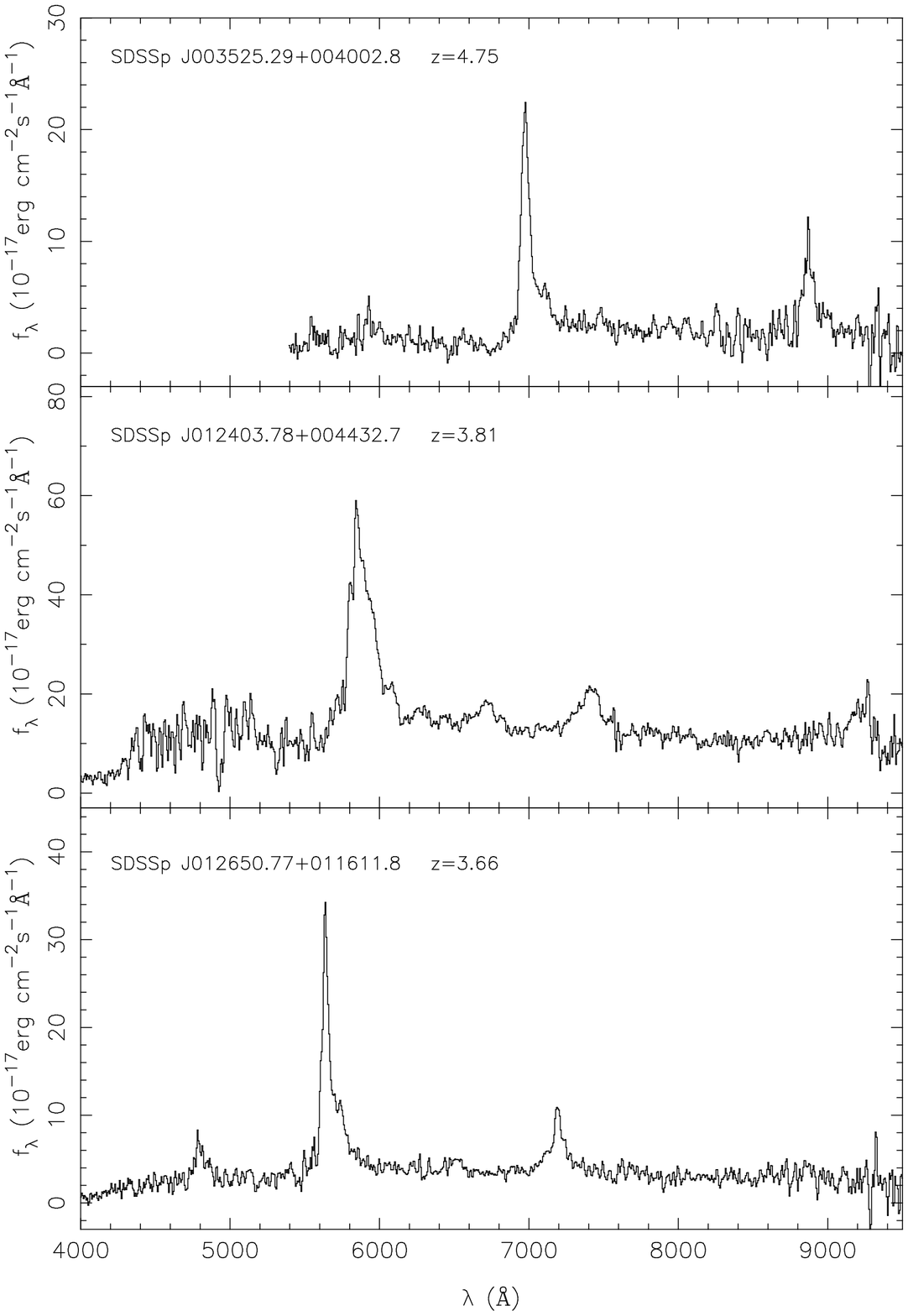}{5.25in}{0.}{65.}{65.}{-207}{-50}
\figurenum{3}
\protect\caption
{\footnotesize
ARC 3.5m/DIS spectra of 15 new SDSS quasars.
The spectral resolution is about 12 \AA\ in the blue and
14 \AA\ in the red.
Each pixel represents 12.4\AA.}
\end{figure*}

\begin{figure*}[t]
\plotfiddle{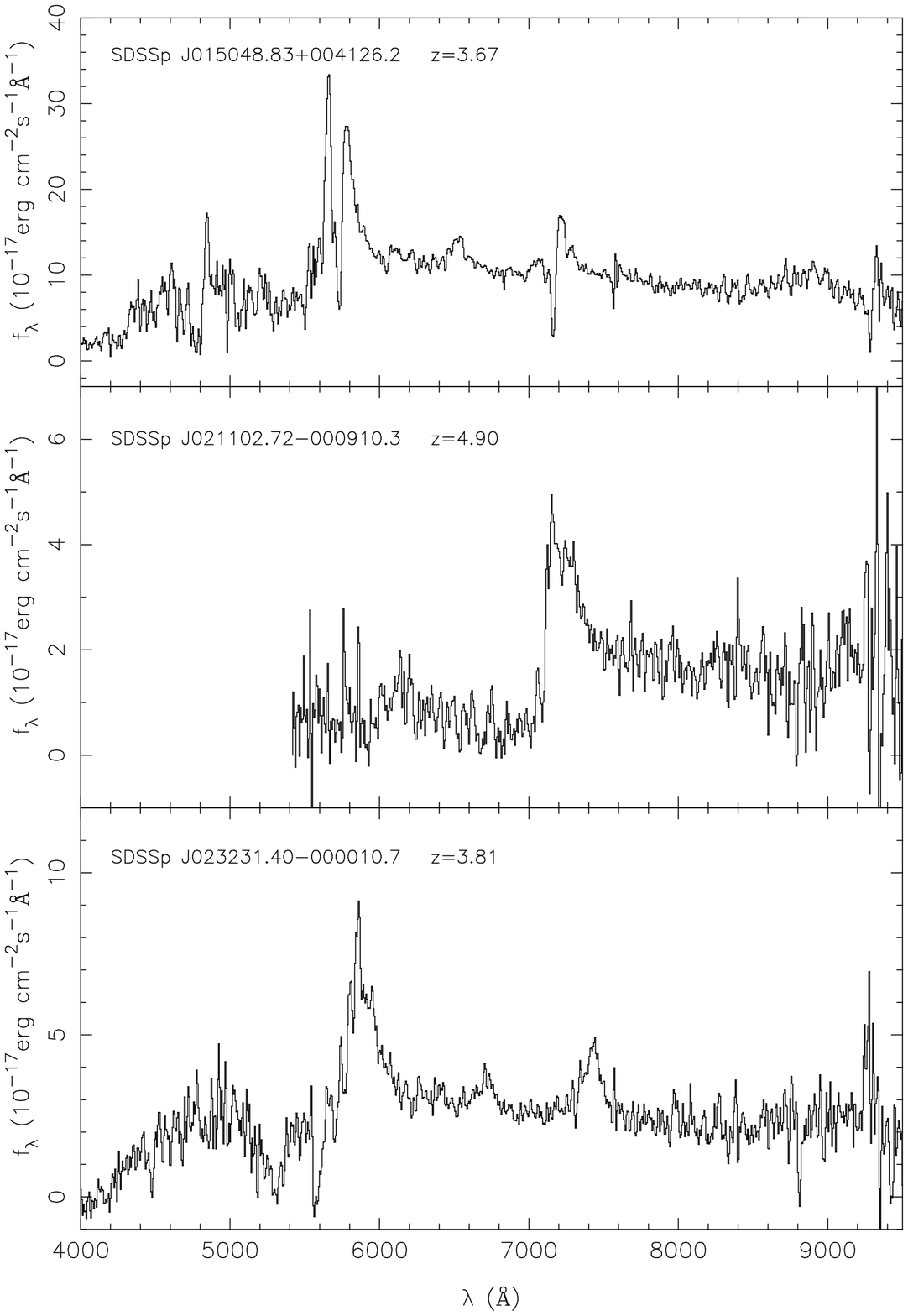}{5.25in}{0.}{65.}{65.}{-207}{-50}
\figurenum{3}
\protect\caption
{\footnotesize
Continued}
\end{figure*}

\begin{figure*}[t]
\plotfiddle{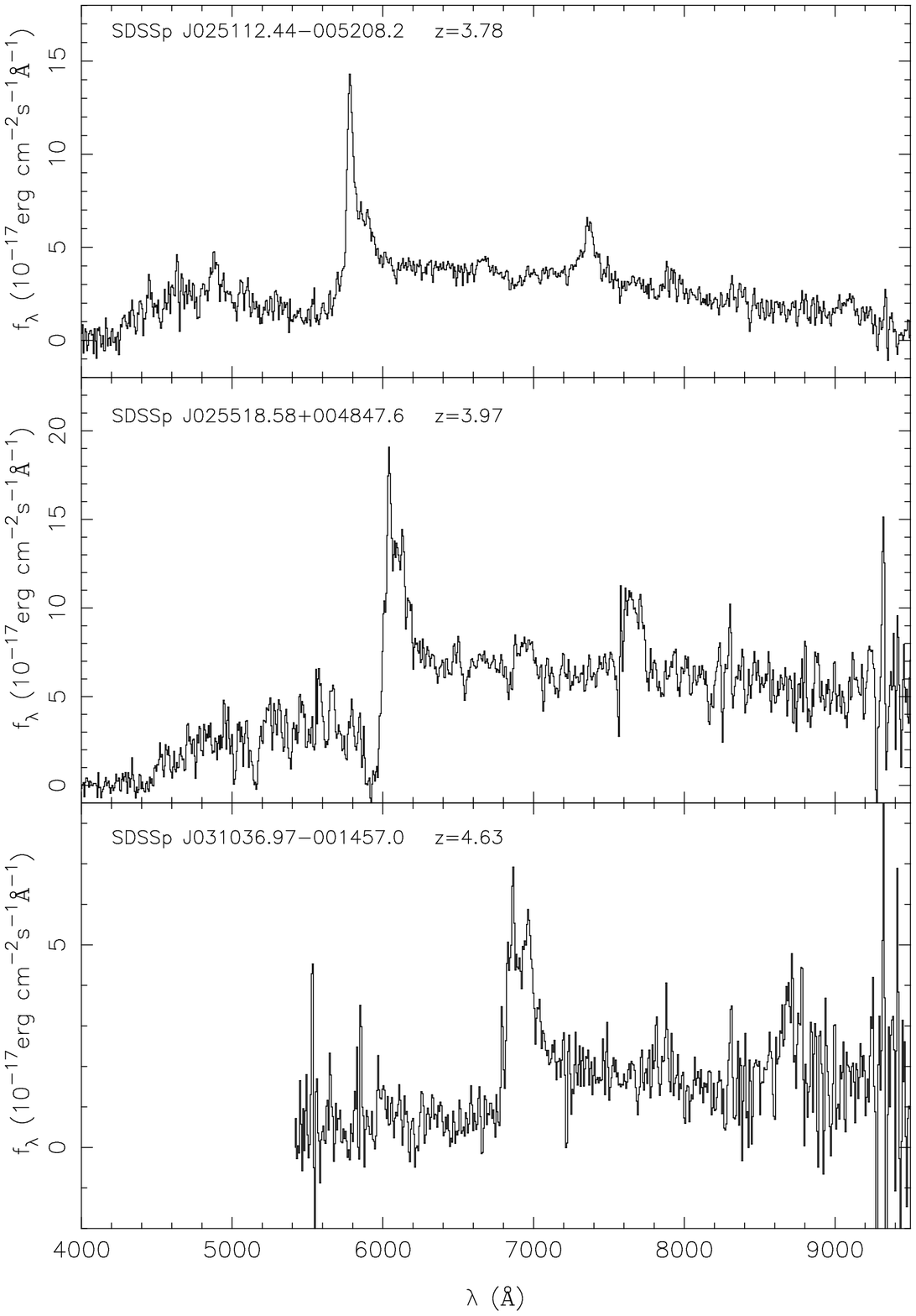}{5.25in}{0.}{65.}{65.}{-207}{-50}
\figurenum{3}
\protect\caption
{\footnotesize
Continued}
\end{figure*}

\begin{figure*}[t]
\plotfiddle{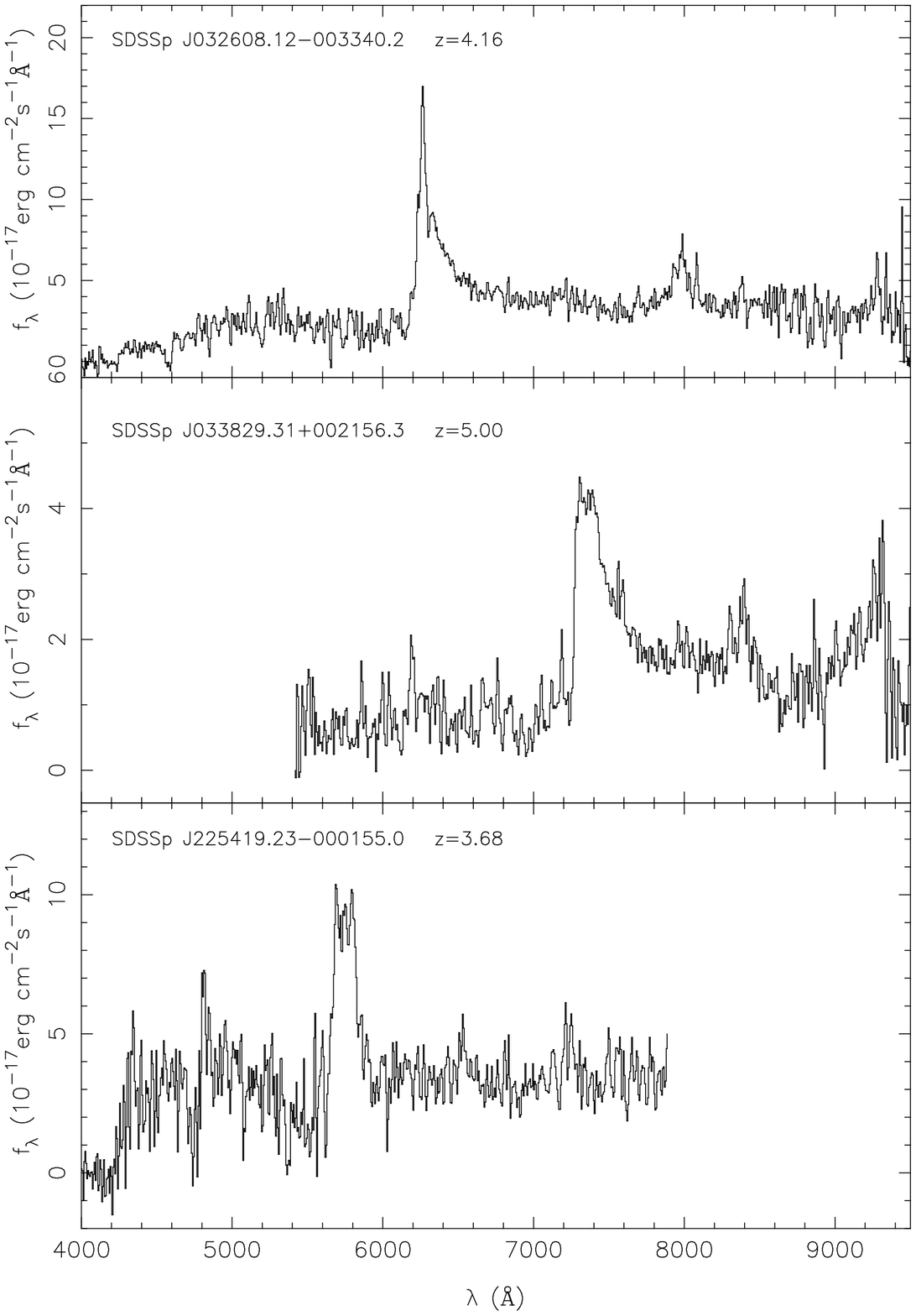}{5.25in}{0.}{65.}{65.}{-207}{-50}
\figurenum{3}
\protect\caption
{\footnotesize
Continued}
\end{figure*}

\begin{figure*}[t]
\plotfiddle{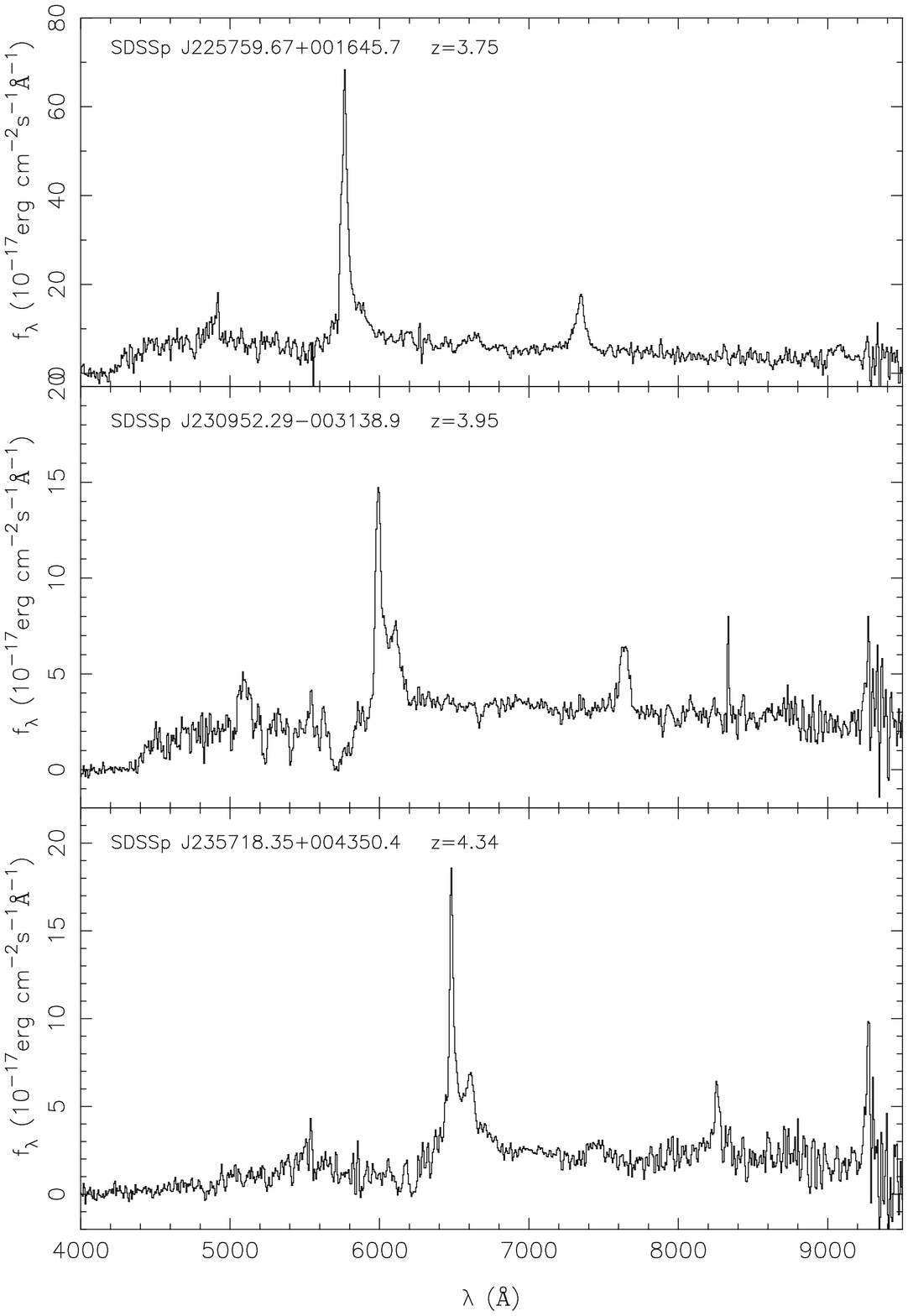}{5.25in}{0.}{65.}{65.}{-207}{-50}
\figurenum{3}
\protect\caption
{\footnotesize
Continued}
\end{figure*}

Of the ten objects not observed, eight are $gri$ candidates, and two
are $riz$ candidates. 

Of the ten candidates whose spectroscopy shows them not to be
high-redshift quasars, five are late type stars. 
They are all faint $riz$ candidates, scattered into the selected region
due to large photometric errors (especially in $z^*$, see \S 6).
The low signal-to-noise ratio of the spectra of the remaining five
candidates does not allow a definitive classification, but their
spectral features are consistent with that of compact galaxies at
$z \sim 0.4$ with a very flat continuum for $\lambda_{rest} > 4000$ \AA.
The broad-band colors inferred from the spectra of these non-quasars
are consistent with the SDSS measurements. 
Some of them may be compact ``E+A'' galaxies (\cite{DG82}, \cite{ZZ96}).
At $z\sim 0.4$, their broad-band colors resemble those of quasars at $z \sim 4$.
The prominent Balmer jumps in ``E+A'' galaxies seen through broad-band
filters resemble the Lyman break in quasars, 
while their flat continua from the ``A'' component in the red resemble
the power-law continua of quasars.  
They are interesting objects in their own right, and will be discussed
in a separate paper (cf., \cite{Fan98}).  

In summary, we have obtained spectra or identifications of 17 of the
25 $gri$ candidates from Run 94. 
Among them, 2 are previously known quasars, 10 are new
SDSS quasars, and 5 are unidentified with spectra consistent
with those of compact galaxies.
Eight of the 10 $riz$ candidates from Run 94 are identified; 3 are new
SDSS quasars at $z=4.75, 4.90$ and 5.00, and 5 are late-type stars.
Two additional new SDSS quasars are also identified.
SDSSp J025518.58+004847.6 (z=3.97) was selected from
Run 94, but does not meet the color selection criteria
(Eq.~2); 
SDSSp J030136.97-001457.0 (z=4.63) was selected
from Run 109.

\section{Spectra of New SDSS Quasars}

Figure 3 gives the low resolution spectra of 
the 15 new SDSS quasars. 
Tables 2 and 3 summarize the spectral properties of these quasars.

Table 2 gives the emission line properties. 
Central wavelengths and rest frame equivalent widths of five major
emission lines, OVI$\lambda$1034, Ly$\alpha$+NV$\lambda$1216+1240,
OI+SiII$\lambda$1306, SiIV+OIV]$\lambda$1402, and CIV$\lambda$1549,
are measured by Gaussian fitting.
In particular, the  Ly$\alpha$+NV blend is
fitted to two Gaussians in all but one case.
The wavelength given in Table 2 refers to that of the Ly$\alpha$
component, while the equivalent width is  the sum of the two.
The emission line properties for these quasars are very similar to
those of other quasars at $z\gtrsim 4$ (\cite{SSG89}, \cite{SSG91a},
\cite{Kennefick95}, \cite{SL96}).
The  properties of the four quasars at $z>4.6$ show  
no obvious difference from those at lower redshifts
(see also \cite{SSG91b}).

The emission line redshift (Table 3) is calculated by taking the weighted mean
of the redshift from OI+SiII, SiIV+OIV] and CIV lines, when they are
detected; Ly$\alpha$ is not used because of the strong absorption on
its blue side. 
For all but two cases, at least two of the three lines are detected.
Two quasars (SDSSp J021102.72$-$000910.3 and SDSSp J230952.29$-$003138.9)
have only CIV detected, and we use its redshift here. 
The uncertainty in redshift determination is dominated by discrepancies of the
redshifts of different lines.
The statistical error from a single line measurement is typically much 
smaller.
The redshift uncertainty quoted in Table 3
is the scatter of the measurements from different lines. 
If the scatter is smaller than 0.01 (or in the cases when only CIV is detected),
we adopt a lower limit of 0.01. 

Table 3 also gives AB$_{1450}$, the AB magnitude of the quasar continuum at
1450\AA\ (rest frame) and $\rm M_{B}$ of each quasar. 
They are corrected for Galactic extinction using the reddening map of
\cite{Schlegel98}. 
We use a cosmology of $q_{0}=0.5$, $h=0.5$, and a power law index of --0.5
for the quasar continuum when calculating $\rm M_{B}$.  The absolute
magnitudes are comparable to that of 3C$\,$273, which is $M_B =
-27$ in this cosmology. 

%
%
Although the quality of our spectra is not high, precluding an
accurate measurement of the flux decrement due to the Ly$\alpha$
forest, more than 70\% of the
continuum  radiation blueward of Ly$\alpha$ emission line is absorbed
by the forest for our highest-redshift quasars. This is quite
consistent with what has been seen with previously known quasars at
similar redshifts. 

Strong Lyman Limit Systems (LLSs) are visible in all quasars that 
have sufficient signal-to-noise ratio in that part of the spectra.
Their corresponding redshifts are listed in Table 3. 
For the four quasars at $z>4.6$, the spectra near the
Lyman Limit are too noisy to decide where the absorption occurs.
In addition, there are several damped Ly$\alpha$ absorption line
candidates (also listed in Table 3); one quasar shows evidence of a
Broad Absorption Line (BAL) system. 

\subsection{Notes on Individual Objects}

\noindent {\bf SDSSp J003525.29+004002.8} ($z=4.75$). This quasar shows very
strong Ly$\alpha$ and CIV lines. 
It has a detectable OI+SiII line, but the SiIV+OIV] line is undetected,
the only such case among the 15 quasars.


\noindent {\bf SDSSp J015048.83+004126.2} ($z=3.67$). This is a mini-BAL quasar
(e.g., \cite{Barlow97}, \cite{Churchill99}), with
strong self-absorption in all detected lines. The CIV
trough has a FWHM of $\rm \sim 4000\ km\ s^{-1}$. 
The LLS at $z_{abs}=3.71$ is probably associated with the BAL.

\noindent {\bf SDSSp J023231.40-000010.7} ($z=3.81$). The strong Ly$\alpha$
absorption feature at 5610\AA\ is at the same redshift as the LLS in
this quasar. 


\noindent {\bf SDSSp J032608.12-003340.2} ($z=4.16$). The spectrum reveals two features
that could be $\tau < 1$ 
LLSs at $z_{abs}=4.15$ and 4.00.  There is a large optical depth LLS
at $z_{abs}=3.67$. 

\noindent {\bf SDSSp J033829.31+002156.3} ($z=5.00$). This is the
quasar with the highest known redshift to date. 
The Ly$\alpha$, SiIV+OIV] and CIV lines are clearly detected, but an
accurate redshift determination is not straightforward.
The Ly$\alpha$ emission line yields a redshift of 5.08, but its blue
wing is strongly affected by absorption.  The other two lines
detected, SiIV+OIV] and CIV, give a consistent redshift of 5.00.
However, the SiIV+OIV] line is quite weak, while the red side of the
CIV line is strongly affected by the strong atmospheric absorption at 9300\AA.
The redshift determination is quite uncertain.
We have adopted a redshift of 5.00$\pm$ 0.04 for this object.


 
\section{Discussion}

The success rate of the quasar search from these SDSS commissioning  
data is substantially higher than that of previous multicolor surveys
at similar magnitude and redshift range 
(e.g. \cite{Kennefick95}, \cite{Hall96}).
Although the current sample is not complete, among the 25 uniformly
selected candidates in Run 94
(including the 2 previously known high-redshift quasars) that have been
identified, 15 are quasars at $z>3.6$, a success rate of 60\%,
while previous surveys have had success rates less than 20\%.
In particular, among the 17 $gri$ candidates that are identified, 
12 are quasars at $z>3.6$, a success rate of 70\%.
This is due to two reasons: (1) the high photometric accuracy
of the SDSS data; and (2) the fact that we have been very conservative in
our candidate selection  (\S 3).   We expect
that as we extend our candidate search to fainter flux levels and
nearer to the stellar locus,  our success rate will decline.

The success rate of $riz$ candidates appears
much lower (3 quasars out of 8 candidates in Run 94, or 38\%). 
The random photometric error in the $z'$ band is 0.15 mag at $z^* \sim
19.8$ for Run 94. 
For quasars at $z \sim 5$, where we must rely on the $z'$ measurements,
a large contamination from red stars is expected. 

Although the SDSS can detect quasars at $z$ up to 7 in the $z'$ band,
selection of quasars at $z>5.5$ (if they exist) may prove quite difficult. 
Figure 1(c) shows that the quasar locus approaches the locus of very
late type stars; this happens for redshifts $z > 5$. 
At even higher redshifts, quasars become undetectable in the $r'$
band, and additional information in the near IR may be required to distinguish
quasars from very red late type stars (such as L-type stars, \cite{Kirk99}).

The survey we have presented here  does not result in a complete sample
for three reasons.  First, 
the photometric calibration of these data is preliminary and
has larger errors than what will be achieved by the SDSS after 
the commissioning period.  This will affect the selection of our
sample at some level, although already our photometric accuracy is
superior to that of most previous multicolor searches for
high-redshift quasars. 
Second, the color cuts applied on the data, although objective, are 
largely arbitrary. They  may miss a considerable fraction of high-redshift
quasars, although we have seen that the quasars we have found follow
the model quasar locus quite closely. 
Third, we have not yet finished observing all our
candidates.  

The current data do not allow us to calculate  the selection function
reliably, so we do not intend to calculate the quasar
number density or  
luminosity function at high-redshift from these quasars at this time. 
However, as a consistency check, we compare the number of quasars observed
to date  with the predicted number counts from Fan (1999) in Table 4.
Fan (1999) predicts the quasar number counts in SDSS bands based
on the quasar luminosity function of \cite{Pei95} for $z<4$ and
Schmidt, Schneider \& Gunn (1995) for $z>4$.
The quasar number counts are based on those identified from
Run 94 (110 deg$^2$).
Since we have not observed all the high-redshift quasar candidates,
nor do we address the problem of completeness, 
the observed  numbers  in Table 4 are only lower limits.
Within the errors from small number statistics, the predicted numbers and
the numbers observed are consistent.
Table 4 shows that although previous surveys have been less efficient
than the present work,
there is no indication that they missed a large fraction of high-redshift
quasars. 

Even though the data in this study have not yet achieved
the SDSS survey standards in imaging quality and in photometric calibrations,
the high quality CCD data and powerful survey software enable 
efficient selection of high-redshift quasars.
The data described in this paper  cover only $\sim$ 1\% of
the SDSS survey area. 
The largest samples that have been used in determining the evolution of the quasar luminosity
function at $z>4$ consist of only $\sim 10$ quasars
(\cite{SSG95}, \cite{Kennefick95}).
In the entire SDSS survey, we expect to find $\sim 600$ quasars
with $z>4$ at $i'<20$, with a significant number of them at $z>5$. 
These quasars will be selected uniformly and will cover a large area of the sky.
This sample will be invaluable in understanding the large scale distribution
and evolution of high-redshift quasars, as well as providing bright quasars
at high redshift for detailed studies of quasar absorption systems.

The Sloan Digital Sky Survey (SDSS) is a joint project of the
University of Chicago, Fermilab, the Institute for Advanced Study, the
Japan Participation Group, The Johns Hopkins University, the
Max-Planck-Institute for Astronomy, Princeton University, the United
States Naval Observatory, and the University of Washington.  Apache
Point Observatory, site of the SDSS, is operated by the Astrophysical
Research Consortium.  Funding for the project has been provided by the
Alfred P. Sloan Foundation, the SDSS member institutions, the National
Aeronautics and Space Administration, the
National Science Foundation, the U.S. Department of Energy, and the
Ministry of Education of Japan.  XF and MAS acknowledge additional
support from Research Corporation, NSF grant AST96-16901, the
Princeton University Research Board, and an Advisory Council
Scholarship. DPS acknowledges the support of NSF grant AST95-09919.
We thank Wolfgang Voges for useful comments on the manuscript.  We
also thank 
the usual expert assistance of Bruce
Gillespie, Karen Gloria, Camron Hastings, Tia Hoyes, and Russet
McMillan.

\section*{Appendix: The Definition of Asinh Magnitudes}

\begin{deluxetable}{crrrrr}
\tablenum{2}
\tablecolumns{6}
\tablewidth{0pc}
\tablecaption{Emission Line Properties of SDSS High-redshift Quasars}
\tablehead
{
quasar           &      OVI  & Ly$\alpha$ & OI+SiII & SiIV+OIV] & CIV   \\
                 & 1034   &  1216+1240 & 1306 & 1402  & 1549
}
\startdata
SDSSp J003525.29+004002.8 & 5955 $\pm$ 5  & 7002 $\pm$ 1  & 7509 $\pm$ 10 &       & 8899 $\pm$ 4  \\
                         & 15.4 $\pm$ 4.2 & 82.4 $\pm$ 5.7 & 16.6 $\pm$ 2.3 &       & 66.4 $\pm$ 5.4 \\
SDSSp J012403.78+004432.7 &       & 5879 $\pm$ 1  & 6317 $\pm$ 7  & 6744 $\pm$ 5 & 7434 $\pm$ 3  \\
                         &       & 74.2 $\pm$ 1.6 & 5.2 $\pm$ 0.6 & 10.6 $\pm$ 0.7 & 24.8 $\pm$ 0.9 \\
SDSSp J012650.77+011611.8 & 4832 $\pm$ 2  & 5661 $\pm$ 1  &       & 6535 $\pm$ 12 & 7221 $\pm$ 4  \\
             & 38.5 $\pm$ 1.3 & 76.0 $\pm$ 7.3 &       & 18.6 $\pm$ 2.3& 50.1 $\pm$ 1.9 \\
SDSSp J015048.83+004126.2 &       & 5707 $\pm$ 2  &       & 6559 $\pm$ 4 & 7226 $\pm$ 4  \\
                         &       & 47.9 $\pm$ 2.6 &       & 7.9 $\pm$ 0.7 & 24.0 $\pm$ 0.9 \\
SDSSp J021102.72--000910.3&       & 7187 $\pm$ 3  &       &       & 9138 $\pm$ 24 \\
                         &       & 60.9 $\pm$ 9.7 &       &       & 39.7 $\pm$ 6.9 \\
SDSSp J023231.40--000010.7&       & 5880 $\pm$ 4  &       & 6740 $\pm$ 9 & 7451 $\pm$ 5  \\
                 &       & 48.6 $\pm$ 5.7 &       & 10.1 $\pm$ 1.5& 23.7 $\pm$ 1.5 \\
SDSSp J025112.44--005208.2& 4916 $\pm$ 4  & 5812 $\pm$ 1  &       & 6697 $\pm$ 15 & 7396 $\pm$ 4  \\
                     & 17.5 $\pm$ 1.5 & 75.1 $\pm$ 5.1 &       & 7.1 $\pm$ 1.7 & 19.2 $\pm$ 1.4 \\
SDSSp J025518.58+004847.6 &       & 6069 $\pm$ 2  & 6520 $\pm$ 12 & 6968 $\pm$ 12 & 7671 $\pm$ 5  \\
                         &       & 31.8 $\pm$ 2.6 & 2.3 $\pm$ 1.0 & 8.4 $\pm$ 1.3 & 26.4 $\pm$ 1.8 \\
SDSSp J031036.97--001457.0 &       & 6886 $\pm$ 3  &       & 7892 $\pm$ 19 & 8727 $\pm$ 10 \\
                         &       & 56.9 $\pm$ 6.7 &       & 19.2 $\pm$ 4.6& 47.7 $\pm$ 5.8 \\
SDSSp J032608.12--003340.2&       & 6292 $\pm$ 1  &       & 6740 $\pm$ 9 & 7451 $\pm$ 5  \\
                         &       & 58.8 $\pm$ 4.6 &       & 6.4 $\pm$ 1.9 & 22.0 $\pm$ 2.2 \\
SDSSp J033829.31+002156.3 &       & 7386 $\pm$ 8  &       & 8419 $\pm$ 12 & 9299 $\pm$ 12 \\
                         &       & 71.5 $\pm$ 14.5&       & 20.3 $\pm$ 3.4& 13.9 $\pm$ 3.0 \\
SDSSp J225419.23--000155.0&       & 5782 $\pm$ 8  &       & 6561 $\pm$ 9 & 7265 $\pm$ 7  \\
                         &       & 87.2 $\pm$ 17.7&       & 8.2 $\pm$ 1.9 & 12.9 $\pm$ 2.2 \\
SDSSp J225759.67+001645.7 & 4925 $\pm$ 4  & 5794 $\pm$ 1  &       & 6657 $\pm$ 11 & 7371 $\pm$ 2  \\
                         & 19.7 $\pm$ 1.6 & 72.3 $\pm$ 6.8 &       & 10.6 $\pm$ 2.1& 36.1 $\pm$ 1.9 \\
SDSSp J230952.29--003138.9& 5125 $\pm$ 1  & 6020 $\pm$ 1  &       &       & 7667 $\pm$ 3  \\
                         & 20.1 $\pm$ 0.7 & 63.8 $\pm$ 2.7 &       &       & 15.5 $\pm$ 0.9 \\
SDSSp J235718.35+004350.4 &       & 6510 $\pm$ 1  &       & 7486 $\pm$ 9 & 8286 $\pm$ 2 7 \\
                         &       & 47.1 $\pm$ 2.4 &       & 12.5 $\pm$ 2.1& 20.2 $\pm$ 1.5
\enddata
\tablenotetext{}{The two entries in each line are the central wavelength and rest frame equivalent width
from the Gaussian fit to the line profile, both measured in
\AA{}ngstroms.}
\end{deluxetable}

\begin{deluxetable}{cccccc}
\tablenum{3}
\tablecolumns{6}
\tablecaption{Continuum Properties of SDSS High-redshift Quasars}
\tablehead{quasar               & redshift   & AB$_{1450}$ &$M_{B}$ &$z_{LLS}$&$z_{abs}^a$}
\startdata
SDSSp J003525.29+004002.8 & 4.75 $\pm$ 0.01 & 19.95   & $-$26.66   &    \\
SDSSp J012403.78+004432.7 & 3.81 $\pm$ 0.02 & 18.07   & $-$28.19   & 3.71 &3.08  \\
SDSSp J012650.77+011611.8 & 3.66 $\pm$ 0.01 & 19.58   & $-$26.62   & 3.57?  \\
SDSSp J015048.83+004126.2 & 3.67 $\pm$ 0.01 & 18.35   & $-$27.75   & 3.71  \\
SDSSp J021102.72--000910.3& 4.90 $\pm$ 0.02 & 20.02   & $-$26.63   &     \\
SDSSp J023231.40--000010.7& 3.81 $\pm$ 0.01 & 19.72   & $-$26.54   & 3.61 &3.61, 3.37 \\
SDSSp J025112.44--005208.2& 3.78 $\pm$ 0.01 & 19.55   & $-$26.70   & 3.67  \\
SDSSp J025518.58+004847.6 & 3.97 $\pm$ 0.02 & 18.66   & $-$27.67   & 3.94 &3.89, 3.26 \\
SDSSp J031036.97--001457.0 & 4.63 $\pm$ 0.01 & 20.06   & $-$26.78   &     \\
SDSSp J032608.12--003340.2& 4.16 $\pm$ 0.02 & 19.19   & $-$27.21   & 4.15  \\
SDSSp J033829.31+002156.3 & 5.00 $\pm$ 0.04 & 20.01   & $-$26.56   &     \\
SDSSp J225419.23--000155.0& 3.68 $\pm$ 0.01 & 19.48   & $-$26.73   & 3.64  \\
SDSSp J225759.67+001645.7 & 3.75 $\pm$ 0.01 & 19.05   & $-$27.19   & 3.63  \\
SDSSp J230952.29--003138.9& 3.95 $\pm$ 0.01 & 19.50   & $-$26.82   & 3.82 &3.73, 3.32 \\
SDSSp J235718.35+004350.4 & 4.34 $\pm$ 0.01 & 19.86   & $-$26.61   & 4.29 &4.14, 4.06
\enddata
\tablenotetext{}{Absolute magnitudes assume $H_0 = 50\rm\, km\, s^{-1}\,Mpc^{-1}$, and
$q_0 = 0.5$.}
\tablenotetext{a}{These are redshifts of damped Ly$\alpha$ lines seen in
the spectra.}
\end{deluxetable}

\begin{deluxetable}{ccc}
\tablenum{4}
\tablecolumns{3}
\tablewidth{0pt}
\tablecaption{Number Counts in 110 deg$^2$: Predicted vs. Observed (Lower Limit from Run 94)}
\tablehead
{
 & predicted & observed
}
\startdata
$ z > 3.6$, $i' < 19$ & 3.0 & 7 \\
$ z > 3.6$, $i' < 20$ & 18.1 & 15 \\
$ z > 4.0$, $i' < 19$ & 1.2 & 2 \\
$ z > 4.0$, $i' < 20$ & 6.3 & 7 \\
$ z > 4.5$, $i' < 20$ & 1.9 & 3 \\
\enddata
\label{tab:LF}
\end{deluxetable}

\begin{deluxetable}{l|lll}
\tablenum{5}
\tablecolumns{4}
\tablewidth{400pt}
\tablecaption{Parameters associated with asinh magnitudes}
\tablehead
{
Filter & $b'$ & $m_0$ & $\mu(0)$
}
\startdata
$u^*$ & 0.46 & 23.40 & 24.24 \\
$g^*$ & 0.53 & 24.22 & 24.91 \\
$r^*$ & 0.60 & 23.98 & 24.53 \\
$i^*$ & 0.70 & 23.51 & 23.89 \\
$z^*$ & 0.55 & 21.83 & 22.47
\enddata
\tablenotetext{}{The softening parameter $b'$ (in units of DN/second),
magnitude zero points $m_0$ corresponding to a count rate of
1 DN per second for SDSS imaging run 94,
and asinh magnitude $\mu(0)$ at zero flux for the preliminary
SDSS color system $u^*$, $g^*$, $r^*$, $i^*$, and $z^*$.
The $b'$ values are fixed for the SDSS, but the $\mu(0)$ values are of course
determined for each run separately.}
\label{tab:asinh}
\end{deluxetable}

Lupton {\em et al.}~(1999a) describe the rationale for a refinement of the usual
logarithmic magnitude scale, to better handle the low signal-to-noise
ratio regime, and the possibility of zero fluxes.  The asinh magnitude
for an object of observed flux $f$
is defined as
\begin{equation}
\mu(f) = \mu(0) -a\, {\rm sinh}^{-1}\left({{f} \over {2\, b'}}\right)
\end{equation}
where $a \equiv 2.5/\ln 10 = 1.08574$, $\mu(0)$ is a normalizing magnitude,
setting the zero-point of the scale, and $b'$ is a softening
parameter.  We set $b'$ equal to the sky noise within a PSF, which
optimizes the error properties of this modified magnitude.  For
signal-to-noise ratios greater than 5, these magnitudes are
essentially identical to standard logarithmic magnitudes.

We have adopted this magnitude scale for the measurements given in
Table 1.  For reference, a detection of zero flux has magnitude
$\mu(0)$, and a formal error
of 0.52 mag in this system, thus a 1-$\sigma$ detection of flux
has an asinh magnitude $\mu(0) - 0.52$. The values of $\mu(0)$ and $b$
are given in Table 5 for each color.

\end{document}